# Hierarchical Low Power Consumption Technique with Location Information for Sensor Networks

Susumu Matsumae
Graduate School of Science and Engineering
Saga University
Saga 840-8502, Japan

Fukuhito Ooshita
Graduate School of Information Science and Technology
Osaka University
Osaka 565-0871, Japan

*Abstract*—In the wireless sensor networks composed of battery-powered sensor nodes, one of the main issues is how to save power consumption at each node. The usual approach to this problem is to activate only necessary nodes (e.g., those nodes which compose a backbone network), and to put other nodes to sleep. One such algorithm using location information is GAF (Geographical Adaptive Fidelity), and the GAF is enhanced to HGAF (Hierarchical Geographical Adaptive Fidelity). In this paper, we show that we can further improve the energy efficiency of HGAF by modifying the manner of dividing sensor-field. We also provide a theoretical bound on this problem.

## I. INTRODUCTION

Wireless sensor networks have gained much attention in recent research and development. In the wireless sensor networks, battery-powered sensor nodes are placed on the observation area, and the sensed data is transmitted to the observer by multi-hop communication between nodes. Traditionally, the routing protocols for these networks have been evaluated in terms of packet loss rates, routing overhead, etc. However, since wireless sensor networks are usually deployed using battery-powered nodes, the optimization of routing protocol's energy consumption is also important [1], [5], [6].

The usual technique for designing an energy-efficient routing protocol is to activate only necessary nodes (e.g., those nodes which compose a backbone network) and to put other nodes to sleep [2], [3], [4], [7], [8]. Among these protocols, in this paper we focus on GAF (Geographical Adaptive Fidelity) [8] and its extended versions called HGAF (Hierarchical Geographical Adaptive Fidelity) and eHGAF (extended HGAF) [4].

In this paper, we show that we can improve the energy efficiency of eHGAF[4] by modifying the manner of dividing sensor field. In the GAF-based algorithms, the sensor field is partitioned by regions called cells, and the cell size affects the energy efficiency of the protocols. Table I summarizes the maximum cell sizes for the GAF-based methods. As shown in Table I, in this paper we successfully obtain the cell size of $\sqrt{3}R^2 \approx 1.73205R^2$, which is 73.205% larger than that of eHGAF [4] whose cell size is $R^2$.

In this paper, we also study an upper bound on the cell size, and prove the upper bound $\pi R^2 - \Delta$ where $R$ is the radio range of each sensor node and $\Delta = \frac{4\pi - 3\sqrt{3}}{6}R^2$. Since this upper bound is approximately $1.91R^2$, our method which attains $\sqrt{3}R^2$ is fairly closed to the theoretical upper bound.

The rest of this paper is organized as follows. Section II explains the outline of GAF, HGAF, and eHGAF. Section III shows that we can further increases the energy efficiency of eHGAF. Section IV provides the upper bound on the cell size. And finally Section V offers concluding remarks.

TABLE I.  MAXIMUM CELL SIZES FOR GAF-BASED PROTOCOLS

|  | *Max cell size* |
|---|---|
| GAF [8] | $\frac{1}{5}R^2$ |
| HGAF [4] | $\frac{1}{2}R^2$ |
| eHGAF [4] | $R^2$ |
| eHGAF with triangle cells (this paper) | $\frac{3\sqrt{3}}{4}R^2$ |
| eHGAF with two cell types (this paper) | $\sqrt{3}R^2$ |

Here, *R* is the radio range of each sensor node.

## II. PRELIMINARIES

Throughout the paper, as in [4], [8], we assume that the radio range of each node is $R$ and is unchanged during the operation, and that every node knows its own location information.

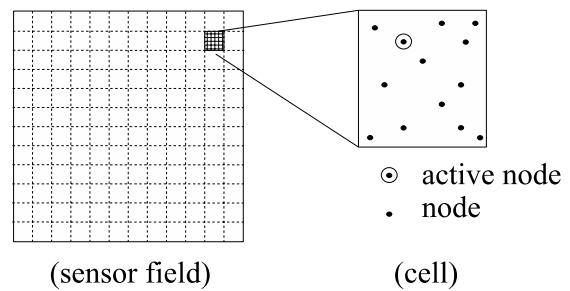

Fig.1. A sensor field divided by square cells

### A. GAF (Geographical Adaptive Fidelity)

In GAF [8], the entire sensor field is divided into virtual sub-fields called *cells*. In each cell, a node called *active node* is chosen. These active nodes have the following two missions:

   *1) The active nodes compose a backbone network for inter-cell data transmissions. Every data across cell-boundaries is conveyed through this back-bone in multi-hop manner.*

   *2) Each active node acts as a gateway node of its own cell. Every transmission across the cell-boundary is via the gateway node.*

Each active node is not fixed, and is properly changed over by the other node in the same cell, according to the remaining amount of battery at the time. The election is dynamically performed by a leader-election algorithm described in [8]. The active nodes are steadily activated, while other nodes are activated only when necessary and are being asleep most of the time.

From the viewpoint of energy consumption, it is preferable to make the cell as large as possible [4]. This is because the larger a cell becomes, the smaller the total number of active nodes in the entire sensor field is. The cell size, however, has an upper bound, and we cannot make it larger without limitation. The upper bound is subject to the communication range of each sensor nodes and the following two requirements:

   (Req. I) Any pair of active nodes can communicate with each other if their cells are adjacent.

   (Req. II) Any active node can communicate with every other node within the cell.

The requirements (Req. I) and (Req. II) are necessary for assuring the missions (I) and (II) of active nodes, respectively.

In GAF, the sensor field is simply divided by square-shaped cells of the same size (see Fig. 1). Let the size of cell be $r \times r$. For GAF, the requirements (Req. I) and (Req. II) are respectively taken on concrete formulas as follows:

$$r^2 + (2r)^2 \leq R^2, \qquad \text{(Req. I-GAF)}$$
$$r^2 + r^2 \leq R^2. \qquad \text{(Req. II-GAF)}$$

The inequality (Req. I-GAF) is due to Fig. 2(a), and the inequality (Req. II-GAF) is due to Fig. 2(b). From these inequalities, we have

$$r \leq \frac{R}{\sqrt{5}}$$

and thus the following claim holds:

**Claim 1** [8] In GAF, the cell size is bounded above by $\frac{R^2}{5}$. ∎

### B. HGAF (Hierarchical Geographical Adaptive Fidelity)

In HGAF [4], the cell size can be $\frac{R^2}{2}$ at largest, by relaxing the dominant condition (Req. I-GAF) of GAF.

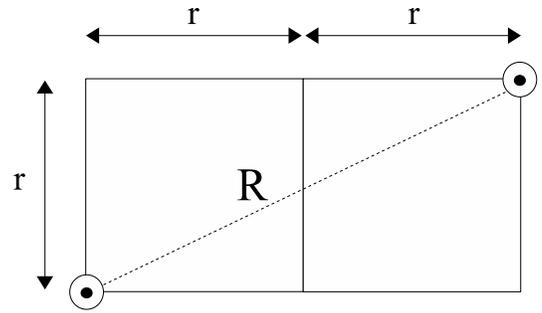

(a) the case where the distance between two active nodes of adjacent cells is the largest.

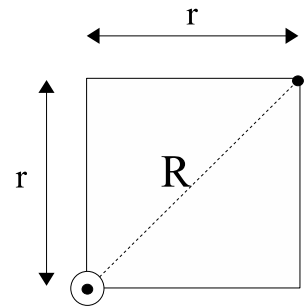

(b) the case where the distance between an active node and a node in the same cell is the largest.

Fig.2.    Examples supporting (Reg. I) and (Req. II) for GAF

The key idea is to avoid the extreme case illustrated in Fig. 2(a). In HGAF, each cell is further divided into smaller squares called *subcells*. A cell of size $r \times r$ is divided into subcells of size $d \times d$. For simplify the exposition, we consider only the case where *r* is divisible by *d*.

A subcell is called *active subcell* if it contains an active node of the cell. In HGAF, active subcells are maintained in the same position of the respective cells, and their positions are synchronously rotated. By this modification, as for (Req. I), we have only to consider the case illustrated in Fig. 3. As for (Req.

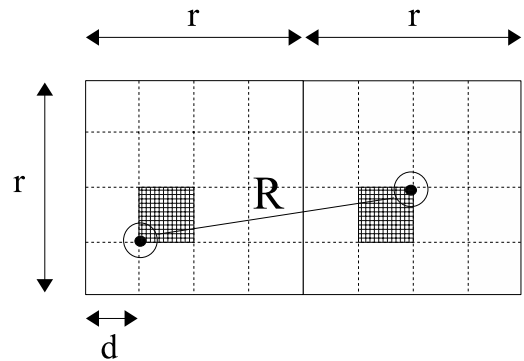

Fig.3.    An example supporting (Reg. I) for HGAF. Here, the distance between two active nodes of adjacent cells is the largest.

II), the inequality is the same as that of GAF. Hence, we have

$$d^2 + (r+d)^2 \leq R^2, \quad \text{(Req. I-HGAF)}$$
$$r^2 + r^2 \leq R^2. \quad \text{(Req. II-HGAF)}$$

From (Req. I-HGAF), we have

$$r \leq \sqrt{R^2 - d^2} - d,$$

and $r$ can be $R$ at largest when we let $d$ be infinitesimal (i.e., the partition of each cell into subcells is infinitely fine-grained). Hence, the constraint (Req. II-HGAF) is the dominant condition here, and thus the following claim holds:

**Claim 2** [4] In HGAF, the cell size is bounded above by $\frac{R^2}{2}$. ∎

*C. eHGAF (extended HGAF)*

In eHGAF [4], the cell size is improved and can be $R^2$ at largest. Here, the dominant constraint for HGAF is relaxed by keeping active subcells centered. For simplicity, we assume that $r$ is divisible by $d$ and that the quotient is an odd number.

To place an active subcell in the center of its cell, the cell-boundaries are synchronously slided properly. By this modification, for (Req. II), we have only to consider the case illustrated in Fig. 4. As for (Req. I), the inequality is the same as that of HGAF. Hence, we have

$$d^2 + (r+d)^2 \leq R^2, \quad \text{(Req. I-eHGAF)}$$
$$2\left(\frac{r+d}{2}\right)^2 \leq R^2. \quad \text{(Req. II-eHGAF)}$$

From (Req. I-eHGAF), $r$ can be $R$ at largest when we let $d$ be infinitesimal. From (Req. II-eHGAF), $r$ can be $\sqrt{2}R$ at largest for infinitesimal $d$. Hence, the constraint (Req. I-eHGAF) is the dominant condition here, and thus the following claim holds:

**Claim 3** [4] In eHGAF, the cell size is bounded above by $R^2$. ∎

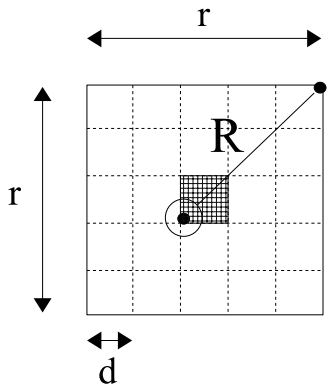

Fig.4. An example supporting (Req. II) for eHGAF. Here, the distance between an active node and a node in the same cell is the largest.

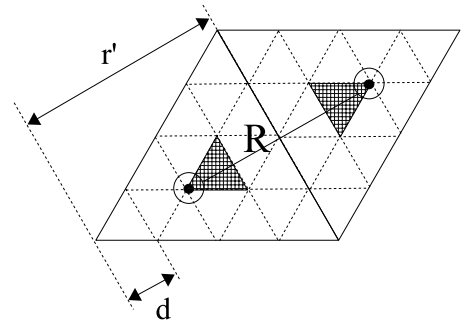

Fig.5. An example supporting (Req. I) for eHGAF with triangle cells. Here, the distance between two active nodes of adjacent cells is the largest.

### III. PROPOSED METHOD

*A. Cell Enlargement by Changing Cell Shape*

In eHGAF with triangle cells, the upper bound on the cell size obtained for the standard eHGAF (i.e., eHGAF with square cells) is further improved. Actually, the cell size can be approximately $1.29904R^2$ at largest, which is 29.904% larger than that of eHGAF [4].

The main idea is to change the base-shape of each cell (and subcell) to triangle cells.

Previously in GAF, HGAF, and eHGAF, the cells are of square-shape. Although partitioning with squares is regular and natural, a plane can be tiled with other regular polygons such as regular triangle and regular hexagon.

In GAF, the cell size can be $\frac{1}{4\sqrt{3}}R^2 \approx 0.14434R^2$, if we use triangle cells, and be $\frac{3\sqrt{3}}{26}R^2 \approx 0.19985R^2$, if we use hexagon cells. That is, for GAF, we cannot improve the upper bound on the cell size even if we adopt triangle/hexagon cells.

However, in eHGAF, the upper bound on the cell size can be improved up to approximately $1.29904R^2$ if we use triangle cells. See Fig. 5. For simplify the exposition, we assume that $r'$ is divisible by $d$ and that the quotient is $(3c+1)$ for some positive integer $c$.

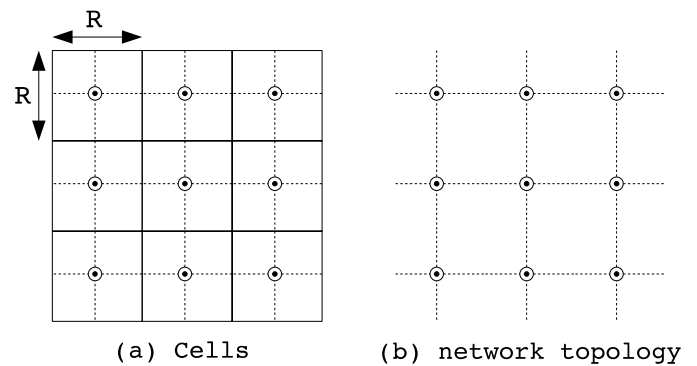

Fig.6. The cell partition, and the network graph of active nodes in eHGAF

This assumption assures that an active subcell can be located in the center (barycenter) of the triangle cell. Here, when $d$ is infinitesimal, we can think that the active subcell can be seen as the point just positioned at the barycenter of the regular triangle. In such a case, it is easy to check that the conditions (Req. I) and (Req. II) become identical, and we have the following one inequality:

$$r' \leq \frac{3}{2}R \quad (d \text{ is infinitesimal}).$$

Since $r'$ is the height of a regular triangle, the maximum size of triangle-shaped cell is calculated as

$$\frac{1}{2} \cdot \frac{2}{\sqrt{3}} r' \cdot r' = \frac{3\sqrt{3}}{4} R^2 \approx 1.29904 R^2.$$

Hence, we obtain the following theorem:

**Theorem 1** In eHGAF with triangle cells, the cell size is bounded above by $\frac{3\sqrt{3}}{4}R^2 \approx 1.29904R^2$. ∎

*B. Cell Enlargement by Reducing Edges*

Next, we show that we can further improve the upper bound on the cell size to $\sqrt{3}R^2 \approx 1.73205R^2$, which is 33.333% larger than that of eHGAF with triangle cells.

In the preceding subsection, the upper bound of $\frac{3\sqrt{3}}{4}R^2 \approx 1.29904R^2$ is obtained by adopting triangle cells. Here in this section, we use a different approach and consider reducing edges of network graph of active nodes.

*0) Relaxing (Req. I):*

In eHGAF with square-cells, the cell size is bounded above by $R \times R$, and each active node is located around the center of its belonging cell. Due to (Req. I), any pair of active nodes must be capable of communicating with each other if their cells are adjacent, and hence the network graph of active nodes becomes a square mesh/lattice. See Fig. 6.

As long as we adhere (Req. I), we cannot break the upper bound of $R \times R$. However, if we loosen (Req. I) properly, we can improve the upper bound further. Here, for example, we consider a version of (Req. I) as follows:

(Req. I') Any pair of active nodes can communicate with each other if their cells are *horizontally* adjacent.

By such a relaxation, though the width of a cell cannot be made longer, the height of it can be $\sqrt{3}R$ at longest. Here, it should be noted that the cell of size $\sqrt{3}R \times R$ can be included in the circle with radius $R$. See Fig. 7.

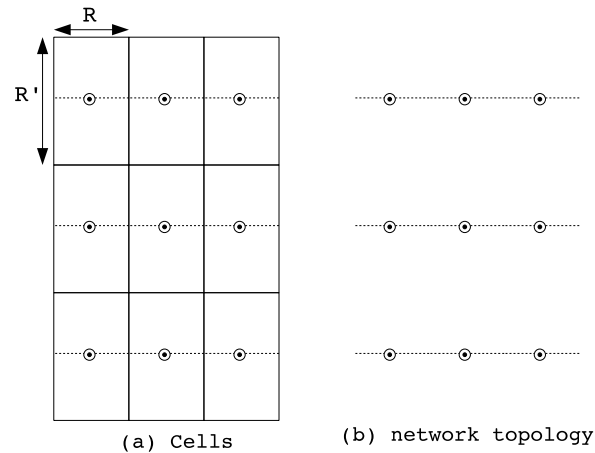

Fig.7. The cell partition of R' × R where R' = √3R, and the network graph of active nodes in eHGAF. Here we adopt (Req. I') instead of (Req. I).

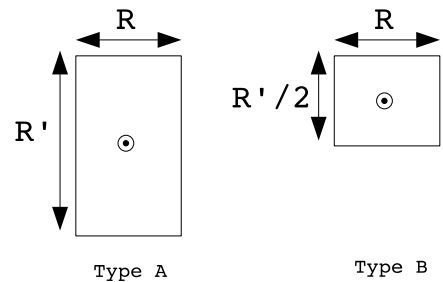

Fig.8. Two types of cells for eHGAF. Here, R' = 3R.

*1) Partition with Two Types of Cells:*

To obtain connectivity of entire network, we consider using another type of cells as well. Here, we use the two types of cells shown in Fig. 8. The height of type B cell is half of that of type A cell. Since the height of type B is $\frac{\sqrt{3}}{2}R$ and is less than $R$, the active node located in the center of it can communicate with its counterpart of upper/lower adjacent cell of type B.

By using both type A and type B cells, we can partition a sensor field in such a way that the network graph of active nodes is connected. See Fig. 9 for an example. If we assume that the columns composed of type B cells are placed every $k$ columns, the average size of cells is calculated as $\frac{\sqrt{3}kR^2}{k+1}$, which converges on $\sqrt{3}R^2 \approx 1.73205R^2$ when $k$ is infinitely large. Here, it should be noted that even when we chose $k = 3$, the average size of cells already becomes $\frac{3\sqrt{3}}{4}R^2$, which is the same size obtained in eHGAF with triangle cells.

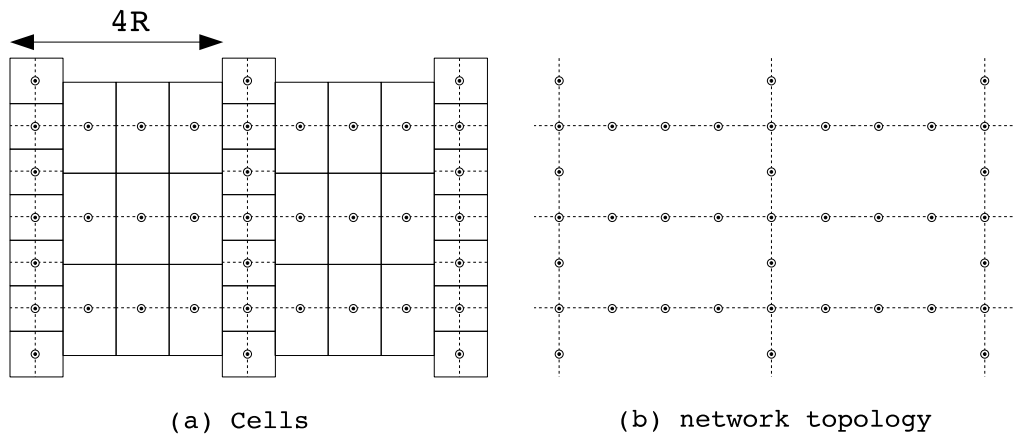

Fig.9. The cell partition with type A and type B cells, and the network graph of active nodes in eHGAF. Here, the columns composed of type B cells are placed every 4 columns.

Hence, we can state the following theorem:

**Theorem 2** In eHGAF with square cells, the cell size can be $\sqrt{3}R^2 \approx 1.73205R^2$, at largest if we permit the existence of active nodes whose degree is less than 4. ∎

## IV. UPPER BOUND OF CELL SIZE IN eHGAF

In this section, we study a theoretical upper bound on the cell size for eHGAF. We show that the cell-size is asymptotically bounded above by $\pi R^2 - \Delta$ in average, where $\Delta = \frac{4\pi - 3\sqrt{3}}{6} R^2$. Here, we do not assume that the shape of sensor field nor that of cells.

To begin with, we introduce the following two propositions.

**Proposition 1** If a sensor field consists of a single cell, the size of entire sensor field can be $\pi R^2$ at largest. ∎

**Proposition 2** If a sensor field consists of 2 cells, the size of entire sensor field can be $2\pi R^2 - \Delta$ at largest. ∎

An example for Proposition 1 is a sensor field whose shape is a circle with radius $R$. An example for Proposition 2 is the one whose shape is the union of two circles such that the radius is $R$ for both circles and that the distance between their centers is $R$.

By generalizing the above two propositions, we can prove the following lemma.

**Lemma 1** If a sensor field consists of n cells, the size of entire sensor field can be $n\pi R^2 - (n-1)\Delta$ at largest. ∎

The Lemma 1 can be proved by mathematical induction on $n$. The base case is due to Proposition 1 and 2. The inductive case can be proved by the following lemma:

**Lemma 2** Let $S_k$ be a sensor field composed of $k$ cells. Construct $S_{k-1}$ from $S_k$ by the following 3 steps:

0) choose any single cell $C$ of $S_k$,
1) remove the region of $C$ if it is covered only by the active node of $C$, and
2) migrate the region of $C$ to the adjacent cell $C'$ if it can be covered by the active node of $C'$.

Then, the following inequality holds:

$$|S_k| - |S_{k-1}| \leq \pi R^2 - \Delta$$

where $|S|$ denotes the area of $S$. ∎

Lemma 2 can be easily checked by the following observation. If an active node of a cell $C_1$ can communicate with its counterpart of adjacent cell $C_2$, then there exists an overlapped area for the circle with radius $R$ whose center is the active node of $C_1$ and that whose center is the active node of $C_2$, and the size of that overlapped area has to be $\Delta$ at least.

Let $P(k)$ denote the following proposition:

If a sensor field consists of k cells, the size of entire sensor field can be $k\pi R^2 - (k-1)\Delta$ at largest.

For the proof of inductive case for Lemma 1, if we assume that $P(k)$ holds and that $P(k+1)$ does not, then we can derive a contradiction by Lemma 2.

From Lemma 1, the average cell size can be calculated as

$$\frac{n\pi R^2 - (n-1)\Delta}{n} = \pi R^2 - \frac{n-1}{n}\Delta.$$

Hence, we can derive the following theorem as follows.

**Theorem 3** In eHGAF, the average cell size can be asymptotically $\pi R^2 - \Delta$ at largest. ∎

## V. CONCLUDING REMARKS

In this paper, we showed the following:

*1) The cell size of eHGAF can be $\left(\frac{3\sqrt{3}}{4}\right)R^2 \approx 1.29904R^2$ at largest if we use triangle cells.*

*2) The cell size of eHGAF can be $\sqrt{3}R \approx 1.73205R^2$ at largest if we permit the existence of active nodes whose degree is less than 4.*

*3) The upper bound on the cell size of eHGAF is $\pi R^2 - \Delta$ where $R$ is the radio range of each sensor node and $\Delta = \frac{4\pi - 3\sqrt{3}}{6}R^2$.*

As shown in Table I, since the previous result [4] can attain only $R^2$ at largest, our results successfully improve the upper bound on the cell-size for the GAF-based methods. Further, since the theoretically obtained upper bound, $\pi R^2 - \Delta$, is approximately $1.91R^2$, our method which attains $\sqrt{3}R^2$ is fairly closed to the theoretical bound.

Since the total number of active nodes in the entire sensor field inversely relate to the cell size, we can say that the energy use of our scheme is more efficient than the previous ones [4], [8]. Table II summarizes the estimated network lifetime for the GAF-based methods. Here, we assume that the network lifetime is proportional to the inverse of the number of active nodes in the entire network.

For future work, we will study whether we can improve the cell-size and/or the theoretical bound further.

TABLE II. NETWORK LIFETIME FOR GAF-BASED PROTOCOLS

|  | *The network lifetime compared to the theoretical upper bound* |
|---|---|
| GAF [8] | 11% |
| HGAF [4] | 26% |
| eHGAF [4] | 52% |
| eHGAF with triangle cells (this paper) | 68% |
| eHGAF with two cell types (this paper) | 91% |
| theoretical upper bound | 100% |

ACKNOWLEDGMENT

This work was partly supported by the JSPS Grant-in-Aid for Young Scientists (B) (23700056).